# On the mechanism of trapped surface formation

D. Valls–Gabaud[1,2] and T. Zannias[3]

*Department of Physics, Queen's University, Kingston, Ontario K7L 3N6, Canada.*

ABSTRACT

By a combination of analytical and numerical methods, the density profile of a momentarily at rest spherical star is varied, and the corresponding response in the area of the spherical shells is monitored. It is shown that the inner apparent horizon (if it exists) must lie within or at most on the star's surface, while no such restriction is found for the outer apparent horizon. However, an apparent horizon lying in the vacuum region will always have non vanishing area, as long as the ADM mass of the system is non zero. Furthermore for density profiles not decreasing inwards, it appears that all spherical trapped surfaces lie on a thick spherical shell. Finally for a uniform density star a simple criterion is found, relating density and proper radius that guarantees the presence or absence of trapped regions.

[1]Institut d'Astrophysique de Paris, 98bis Bld. Arago, 75014 Paris, France.

[2]NSERC International Fellow.

[3]CITA National Fellow.

There has been lately considerable renewed effort aiming to discover conditions, upon a given initial data set, that permit us to infer the presence or absence of trapped surfaces [1,3,4–6]. Although most of the cited work is focused on characterizing the state of initial data triggering the formation of trapped surfaces, it would also be beneficial to understand the mechanism that leads to their formation as well as any other relevant properties. For instance how do spherical trapped surfaces appear and how are they distributed around the centre of the star? Where (if it exists) is the location of the outer apparent horizon? What is the response of the apparent horizon under smooth variations of the parameters describing the initial data?

It is the pupose of the present note to provide, at least for a special class of configurations, answers to the above-mentioned questions. For that we shall consider a sequence of instantaneous states describing spherical stars of proper radius $R$ and proper density $\rho$ just at the onset of their gravitational collapse. By examining the proper area of spheres around the centre of the star as a function of distance away from the the centre, the trapped region can be identified: it lies between succesive maxima and minima of the area function. We vary the density profile (in essence we are moving from one member of the sequence to another) and monitor the behaviour of the area function. Such "variations" of the density cause the appearance or dissappearance of trapped surfaces and forces the inner and outer apparent horizons to "move" around in the star's interior or exterior regions.

Recall that initial data on a time-symmetric slice $\Sigma$ satify the hamiltonian constraint, *i.e.*,

$$^{3}\mathcal{R} \;=\; 16\pi\frac{G}{c^2}\rho \tag{1}$$

where $\rho$ stands for the non-negative density. The assumption of spherical symmetry allows the introduction of geodesic-type coordinates and thus the line element of $\Sigma$ takes the form

$$ds^2 \;=\; dr^2 + B(r)\left(d\theta^2 + \sin^2\theta d\phi^2\right) \qquad 0 \le r \le \infty \tag{2}$$

Evaluating the scalar curvature $^{3}\mathcal{R}$ of the metric (2), equation (1) reduces to

$$-\frac{2}{B}\frac{d^2B}{dr^2} + \frac{1}{2B^2}\left(\frac{dB}{dr}\right)^2 + \frac{2}{B} \;=\; 16\pi\frac{G}{c^2}\rho \tag{3}$$

We look for solutions admitting regular geometry at $r = 0$. This demands

$$B(r)\Big|_{r=0} \;=\; 0 \qquad\qquad \frac{dB}{dr}\Big|_{r=0} \;=\; 0 \tag{4}$$

For later use note that the amount of mass-energy within each sphere of radius $r$ is given by [2]

$$m(r) \;=\; \frac{1}{2}\frac{c^2}{G}B^{1/2}\left[1 - \frac{1}{4B}\left(\frac{dB}{dr}\right)^2\right] \tag{4a}$$

and it obeys, by virtue of Eq. (3) to

$$\frac{dm(r)}{dr} \;=\; 2\pi\rho B^{1/2}\frac{dB}{dr} \tag{4b}$$

Our intention is to investigate the behaviour of the solutions of Eq. (3) subject to (4). A number of conclusions can be drawn by inspection of Eq. (3). At first for $\rho = 0$ a solution obeying (4) is given by $B(r) = r^2$, which corresponds to a flat-three space. Further if a solution $B = B(r)$ admits inflexion points of non zero area (and thus the area function passes through a saddle point) they must be either interior points or lie at the star's surface. At such points, one deduces from Eq. (3) that the density $\rho$ and the area $B(r)$ obey the relation $B\rho = c^2/8\pi G$.

Let us now consider a solution extending from the interior to the exterior vacuum region. Absence of surface layers at the star's surface requires $B(r)$ to be $\mathcal{C}^1$ at $r = R$, *i.e.*, $B(r)$ and $dB/dr$ should be continuous across the surface (note however that $d^2B/dr^2$ may exhibit discontinuous behaviour at $r = R$). Since on the other hand in the vacuum Eq. (3) implies $d^2B/dr^2 > 0$, therefore if the surface of the star is not trapped (*i.e.*, $dB/dr > 0$ there), then the exterior solution will be a monotonically increasing function, that is, trapped surfaces can never develop in the vacuum region. In the opposite case, *i.e.*, if the surface is trapped, Eq. (3) implies that only a local minimum of $B(r)$ can develop in the vacuum region. This outermost local minimum is the outer apparent horizon. Note however that in general it may also lie in the interior. In the case where it lies in the vacuum one may naturally ask: can it have zero area? We shall show that as long as the data possess a non zero ADM mass, this is not possible. In fact evaluating the right hand side of Eq. (4a) on the outer horizon (of zero area), and using Eq. (3) we get :

$$m(r) = \frac{1}{2}\frac{c^2}{G}B^{1/2}\left(2 - \frac{d^2B}{dr^2}\right) = 0$$

which, in view of Eq. (4b) is impossible. This argument also rules out vacuum inflexion points of zero area.

Let us now shift attention to the behaviour of the local maxima of $B(r)$. The first local maximum of $B(r)$ (if it exists) marks the location of the inner apparent horizon. Generically it lies interior to the surface, but one may ask whether the inner horizon lies at the surface of the star. The answer depends on the behaviour of the density profile at the surface. If $\rho(r)$ is continuous at $r = R$ then Eq. (3) implies that the inner horizon cannot be located at the surface. If it exists it must be an interior point. In the case where $\rho(r)$ is discontinuous at $r = R$, Eq. (3) implies that the left derivative $(d^2B/dr^2)_L$ and surface density satisfy

$$\frac{1}{B} = 8\pi\frac{G}{c^2}\rho + \frac{1}{B}\left(\frac{d^2B}{dr^2}\right)_L \tag{5}$$

Therefore arbitrary density profiles allow the possibility that the left second derivative be negative while simultaneously the right hand side of (5) is positive, implying that the surface of the star may become the location of the inner horizon. Notice in that case the $\mathcal{C}^1$ matching of $B(r)$ across the surface dictates that the second right derivative is given by

$$\frac{1}{B}\left(\frac{d^2B}{dr^2}\right)_R = 8\pi\frac{G}{c^2}\rho + \frac{1}{B}\left(\frac{d^2B}{dr^2}\right)_L > 0 \tag{6}$$

implying that the surface must be an outer apparent horizon as sensed by the exterior geometry (this situation will be verified below by an explicit exact equation). In summary, no inner horizon

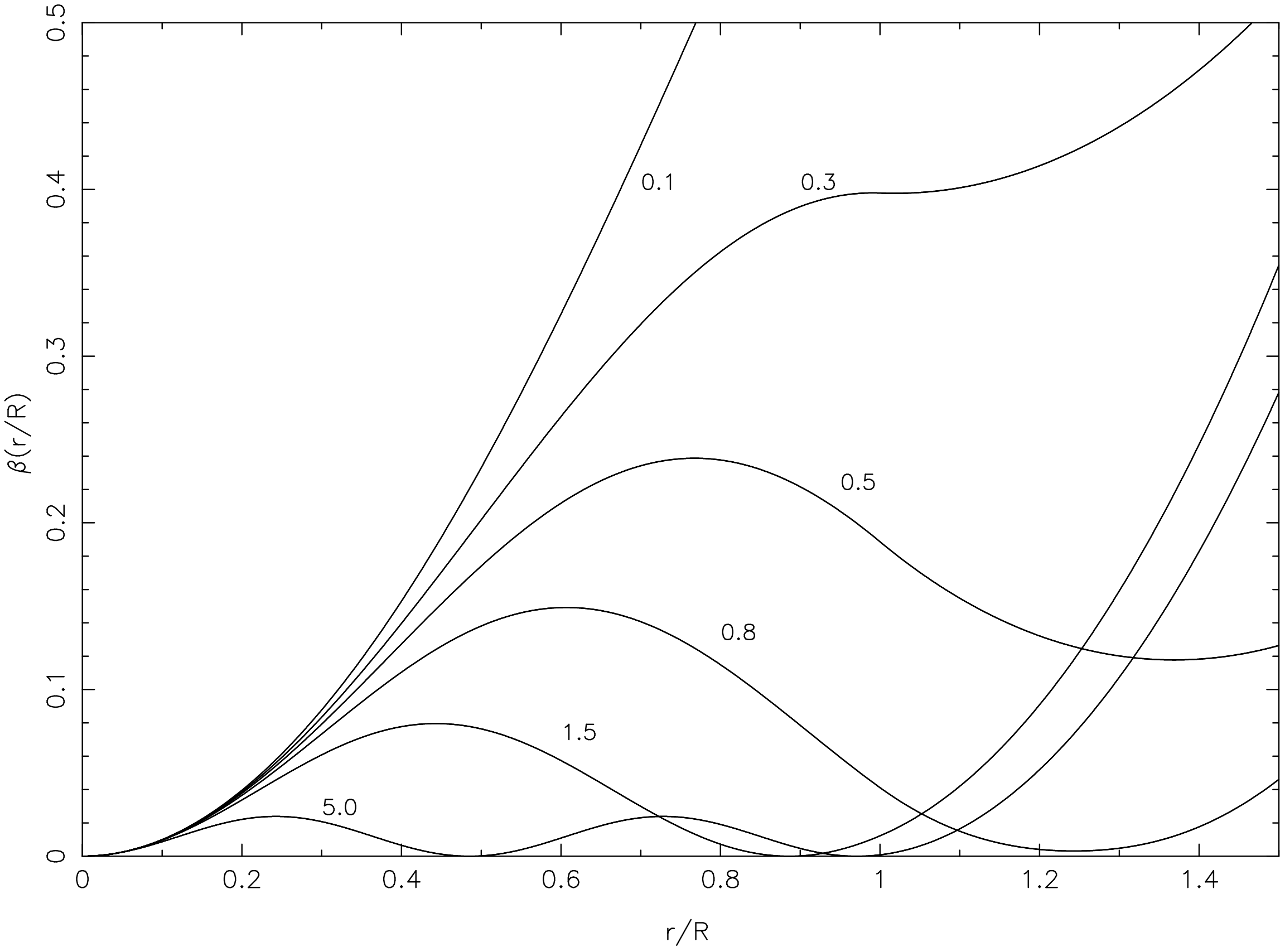

Figure 1: Dimensionless area $\beta$ as a function of the dimensionless proper distance $y = r/R$ for a uniform density distribution. The values of the parameter $\Upsilon_o$ are indicated.

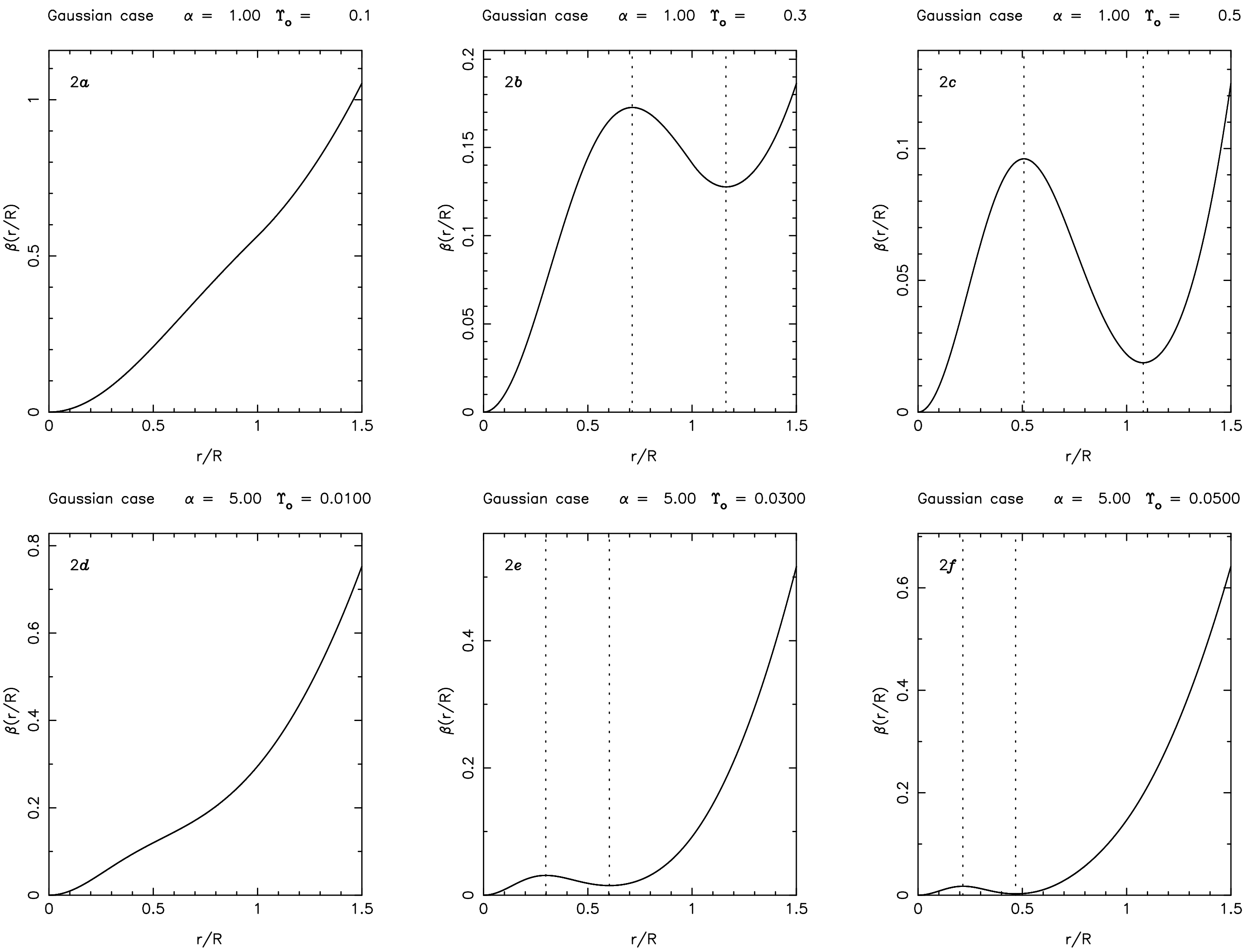

Figure 2: As Figure 1, for gaussian density distributions characterized by $\Upsilon_o$ and $\alpha$. The dotted vertical lines mark the positions of local extrema. Note the presence of a maximum and a minimum (indicative of a trapped surface) in the area above a given threshold in the parameter $\Upsilon_o$, for a given compactness parameter $\alpha$.

can lie exterior to the star in contradistinction to the outer apparent horizon.

How many distinct local maxima and minima are admitted by an arbitrary solution of Eqs. (3) and (4) matched in a $\mathcal{C}^1$ fashion across the surface? The answer to this question is not obvious. The non-linear structure of Eq. (3) does not allow us to get further insight into the nature of the solutions and thus we have to search for exact solutions. For arbitrary density profiles $\rho(r)$ exact solutions appear difficult to find, therefore we shall use numerical techniques to construct solutions of the system. Note that one case where the system is soluble corresponds to uniform density stars, *i.e.*, $\rho = const.$ [7]. Before we comment on this solution, let us rewrite the system in a slightly different form, more suitable for numerical computations. It is convenient to introduce the dimensionless variables

$$y = r/R \qquad \beta = B/R^2 \tag{7}$$

where $R$ is the physical (proper) radius of the star, so that Eqs. (3) and (4) take the dimensionless form

$$-\frac{2}{\beta}\frac{d^2\beta}{dy^2} + \frac{1}{2\beta^2}\left(\frac{d\beta}{dy}\right)^2 + \frac{2}{\beta} \;=\; 16\pi\Upsilon \tag{8}$$

$$\beta\Big|_{y=0} = \frac{d\beta}{dy}\Big|_{y=0} \;=\; 0 \tag{9}$$

where

$$\Upsilon(y) \;=\; \frac{G}{c^2}R^2\rho(r) \tag{10}$$

In this notation, we shall examine two density profiles corresponding respectively to a uniform configuration

$$\Upsilon_U(y) \;=\; \begin{cases} \Upsilon_o & 0 \le y \le 1 \\ 0 & y > 1 \end{cases} \tag{11}$$

and a gaussian one

$$\Upsilon_G(y) \;=\; \begin{cases} \Upsilon_o\, e^{\alpha(1-y^2)} & 0 \le y \le 1 \\ 0 & y > 1 \end{cases} \tag{12}$$

where $\Upsilon_o$ and the compactness parameter $\alpha$ characterize the profiles. The solution of Eqs. (8) and (9) with source given by (11) has the form in the interior ($y \le 1$)

$$\beta(y) \;=\; \frac{3}{8\pi\Upsilon_o}\sin^2\left[\left(\frac{8\pi\Upsilon_o}{3}\right)^{1/2} y\right] \tag{13}$$

In the exterior region Eq. (8) is numerically integrated with boundary conditions dictated by the $\mathcal{C}^1$ matching of the solution on the surface. In Figure 1 solution curves for various values of the dimensionless parameter $\Upsilon_o$ are shown. It is worth to mention briefly a few features of the solution relevant to our discussion. One may easily verify that as long as $\Upsilon_o < 3\pi/32$, $\beta(y)$ increases monotonically. At precisely this critical value $\Upsilon_o = 3\pi/32$, $\beta(y)$ develops the first local maximum taking place at the surface of the star. Note that because of the discontinuity of the density across the surface, this critical point is a local minimum according to the exterior geometry. Thus for this particular value of $\Upsilon_o$ there exists only one trapped surface which is simultaneously the location of the inner and outer apparent horizons. As $\Upsilon_o$ keeps increasing, the degeneracy is lifted and the inner and outer apparent horizons start "moving" on opposite directions (see also Fig. 1). Upon further increase, the inner horizon moves moderately towards the centre while the outer horizon moves towards the surface and simultaneously keeps shrinking (*i.e.*its proper area decreases). Finally at $\Upsilon_o = 3\pi/8$ the outer horizon has zero proper area and resides exactly on the surface of the star. Any subsequent increase of $\Upsilon_o$ causes the outer horizon to move inwards while maintaining its zero area. For even larger values of $\Upsilon_o$ more critical points start to appear (see also Fig. 3a), but such data may be regarded as unphysical since the gravitational field generated by them is so strong that it disconnects the star from the asympotically flat region (for further relevant comments on this issue see ref. [8]). In summary, whenever data obey $3\pi/32 < \Upsilon_o < 3\pi/8$, they admit a distinct inner and outer apparent horizons.

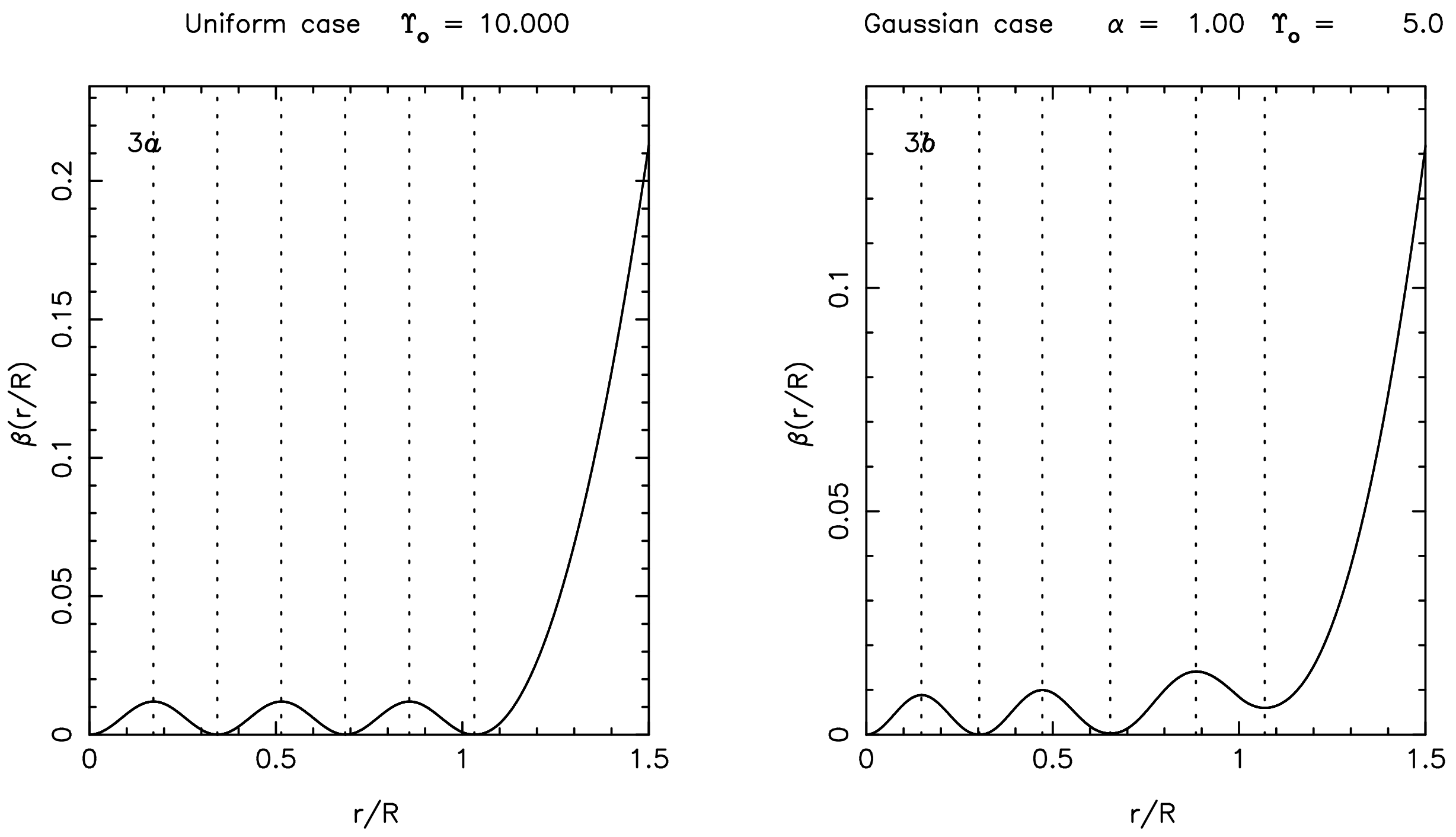

Figure 3: Examples of configurations for which several maxima and minima are present in the dimensionless area $\beta$. *(a)* Uniform density configuration. *(b)* Gaussian density profile.

Solutions of Eqs. (8) and (9) with a gaussian density profile (12) exhibit the same qualitative features as those of the uniform case. In Figure 2 (a-f) we present a few solutions for various values of the parameters $\alpha$ and $\Upsilon_o$. Note however that in this case there exists data that allow both inner and outer apparent horizons of non zero area lying within the interior of the star (Fig. 2e), unlike the solutions for uniform density stars. From the analysis of the solutions it appears a common property worth emphasizing. The system of Eqs. (3) and (4) admits no solutions where two or more consecutive minima (excluding the origin) have values different from zero. In turn, this property implies that a system must first disconnect itself from the asymptotically flat region before a second minimum appears (see Fig. 3 a-b). We like to think that this property of Eqs. (3) and (4) is a generic feature of all solutions, but we have been unable to show this analytically. We discuss elsewhere [9] the impact of these constraints on models of compact astrophysical objects.

Finally it is of interest to know how much of the above described picture is maintained if one deals with configurations lacking time symmetry or spherical symmetry. Although one may anticipate a similar overall behaviour (at least for spherical systems), we ought to bear in mind that there are additional factors entering the problem which may alter some of the features discussed here. We hope to come back to this point elsewhere.

**References**

[1] J.B. Hartle and D.C. Wilkins, Phys. Rev. Lett. **31**, 60 (1973).
P.N. Demmie and A.I. Janis, J. Math. Phys. **14**, 973 (1973).
R. Schoen and S.T. Yau, Commun. Math. Phys. **90**, 575 (1973).
N.O. Murchadha, Phys. Rev. Lett. **57**, 2466 (1986).

[2] This expression for the mass is the familiar Schwarzschild mass written in our coordinate system and taking into account the momentarily static nature of the configuration.

[3] E. Flanagan, Phys. Rev. **D44**, 2409 (1991).
E. Flanagan, Phys. Rev. **D46**, 1429 (1992).
E. Malec, Phys. Rev. Lett. **69**, 946 (1991). and Mod. Phys. Lett. A **7**, 1679 (1992).

[4] P. Bizon, E. Malec and N.O. Murchadha, Phys. Rev. Lett. **61**, 1147 (1988)., Class. Quantum Grav. **6**, 691 (1989)., Class. Quantum Grav. **7**, 1958 (1990).

[5] T. Zannias, Phys. Rev. **D47**, 1448 (1993).

[6] P. Bizon and E. Malec, Phys. Rev. **D40**, 2559 (1989).
R. Beig and N.O. Murchadha, Phys. Rev. Lett. **66**, 2421 (1991).
T. Zannias, Phys. Rev. **D45**, 2998 (1992).
S. Hayward, Class. Quantum Grav. **9**, L115 (1992).

[7] N.O. Murchadha in Ref. [1], P.Bizon, E. Malec and N.O. Murchadha in Ref. [4]. Also the solution can de deduced from the exact four dimensional constant density solution, or the Oppenheimer-Snyder model.

[8] B.J. Carr and S.W. Hawking, Mon. Not. R. Astr. Soc. **168**, 399 (1974).

[9] D. Valls–Gabaud and T. Zannias, unpublished.